\documentstyle[prd,aps,twocolumn,floats]{revtex}
\begin{document}
\twocolumn[\hsize\textwidth\columnwidth\hsize\csname
@twocolumnfalse\endcsname
\preprint{}
\title{Cosmic evolution during primordial black hole evaporation}
\author{Winfried Zimdahl\footnote{Electronic address:
winfried.zimdahl@uni-konstanz.de}}
\address{Fakult\"at f\"ur Physik, Universit\"at Konstanz, PF 5560 M678, 
D-78457 Konstanz, Germany
}
\author{Diego Pav\'{o}n\footnote{Electronic address:  
diego@ulises.uab.es}}
\address{Departamento de F\'{\i}sica
Universidad Aut\'{o}noma de Barcelona,
08193 Bellaterra (Barcelona), Spain}
\date{\today}
\maketitle
PACS numbers {98.80Hw, 04.70Dy, 05.70.Ln}

\begin{abstract}

Primordial black holes with a narrow mass range are regarded as a
nonrelativistic fluid component with an equation of state for dust.
The impact of the black hole evaporation on the dynamics of the early
universe is studied by resorting to a two-fluid model.
We find periods of intense radiation reheating in the initial and  
final stages of the evaporation.
\end{abstract}
\vspace{1.5cm}
]

\section{Introduction}
The interest in primordial black holes (PBHs)
dates back to Zel'dovich and Novikov \cite{ZEL}, Hawking  
\cite{HAW1}, and Carr and Hawking \cite{CHW},
who concluded they should be produced by density
fluctuations in the primordial cosmic fluid, leading
to a wide mass spectrum for these objects. The earlier their
formation, the lower their mass. Since then there has been
a manifold  of proposals for the production of PBHs:
Inhomogeneities triggered in the inflationary period of
cosmic expansion \cite{CL}, first order 
phase transitions \cite{KP,KoSaSa}, 
bubble wall collisions \cite{BCKL}, 
the Gross-Perry-Yaffe mechanism of quantum
gravitational tunneling from hot
radiation \cite{GPY} (see also \cite{KAP} and  \cite{PW}),
quantum pair creation (based on the nonboundary
proposal of the wave function of the universe) during inflation
\cite{RBH}, quantum fluctuations in hybrid inflation
\cite{BLW} (see also \cite{IVAN}), and the decay of cosmic loops
\cite{PZ}, to mention just a few.

This interest in PBHs is understandable as they may
have had a notable impact
on several areas of current interest, such as
baryogenesis \cite{JDB}, dark matter
\cite{BJC}, galactic nucleation \cite{BJC1}, and the
reionization of the universe \cite{MG}.
In addition they can constrain the spectral index of primordial  
density perturbations \cite{GL,GLR}.

A natural outcome of the copious production of black
holes in some of the mentioned scenarios  is that,
sooner or later, a point will be reached  where 
the energy density of the PBH component will significantly contribute
to the energy density of the cosmic medium. 
For example, phase transitions based on grand unified theories 
\cite{KoSaSa} or bubble wall collisions in extended inflation 
\cite{BCKL} may lead to a PBH dominated universe. 
Also, in scenarios of hybrid inflation there may exist a regime 
in which the black holes will dominate the total energy density 
soon after the end of inflation \cite{BLW}. 
This situation may last for some time
before the contribution of the latter component to the total
energy density becomes negligible because of its
evaporation via Hawking's process. (This stage has been termed
``the binary phase" and studied in detail \cite{GD} for
PBHs produced by the Gross-Perry-Yaffe mechanism.)

PBH formation processes from phase transitions, e.g. in connection  
with extended inflation, are characterized by a very narrow mass  
spectrum
(see, e.g., \cite{BCL2} and references therein).
In such a situation the assumption of only one mass
$m _{_{\left(BH \right)}}$ for all members of the population may be  
considered a reasonable approximation. On the other hand,
a black hole of mass $m _{_{\left(BH \right)}}$ may  
thermodynamically be characterized by a temperature
$T _{_{\left(BH \right)}} \propto m _{_{\left(BH \right)}}^{-1}$  
\cite{HAW2}.
Consequently, with a single mass population of PBHs one may  
associate a single temperature as well.
This suggests a picture of the PBH component in which
$T _{_{\left(BH \right)}}$ in some respect plays the role of an  
equilibrium fluid temperature.

On the basis of this interpretation it is the
target of the present paper to study the cosmological dynamics  
during PBH evaporation within a model of two interacting and  
reacting fluids.
The evaporation process in this model is pictured as a decay of the  
PBH ``fluid'' into a conventional fluid.
We discuss the reheating of the latter as a consequence of the  
evaporation and show that one may describe this process in terms of  
a bulk pressure of the cosmic medium as a whole.
The general formalism is then applied to the dynamics of a
Friedmann-Lema\^{\i}tre-Robertson-Walker (FLRW) universeand the back 
reaction of the PBH  
evaporation on the cosmic scale factor is investigated.

To establish a transparent picture we introduce the following  
simplifying assumptions from the outset:
(i) the PBHs can be
thought of as particles of a perfect fluid with the equation
of state for dust,
(ii) both fluids share the same four-velocity,
(iii) all the PBHs start
evaporating at the same time.
Once these assumptions are accepted one
can apply the formalism of interacting and reacting fluids
to the mixture of both components (say radiation
and the PBH fluid)\cite{WZ1}.
Notice that if the PBHs did not
evaporate,  the evolution of the scale factor and the
energy densities of both components would  directly
be given by the well-known results of Jacobs
for a mixture of relativistic particles and
dust \cite{KCJ}. However the fact
that the PBHs radiate makes a big difference
with respect to the rather traditional case of a
mixture of radiation and dust with conserved fluid particle numbers. 

A closely related
work of Barrow {\em et al.} \cite{BCL} considers
the evaporation of PBHs with an initial power-law mass spectrum.
The
black holes in that paper do not constitute, however, a fluid in  
the sense discussed here (such kind of fluid picture looks less  
natural
for a nonsingular distribution
of masses).
In general, the dynamical behavior of both configurations is  
different,
especially with respect to the existence of solutions for which the  
energy density in PBHs is in equilibrium with radiation.

This paper is organized as follows. Section II recalls the general
hydrodynamical formalism for two interacting and reacting perfect
fluids \cite{WZ1,WD1} which previously has been
applied to the reheating phase at the end of the inflationary
expansion \cite{WD2,WDR}. Section III specifies
this framework to a mixture of thermal radiation plus
primordial black holes and derives a general formula for the  
corresponding
entropy production density, which is then used to discuss specific  
features of the evaporation process, especially a ``reheating'' of  
the radiation.
The evolution equations
for the energy densities of each fluid coupled
to the Friedmann equation are solved numerically.
Section IV introduces an
effective one-component model of the cosmic medium and describes  
the PBH evaporation process in terms of an effective bulk viscous  
pressure of the system as a whole. We present a general formula for  
the latter, which is  evaluated explicitly in
section V for the initial stages of the evaporation process.  
Finally, section VI summarizes
the results of this work.
Units have been chosen so that
$c = k_{B} = \hbar = G = 1$.
\section{The general two-fluid model}
The energy-momentum tensor $T ^{ik}$ of the cosmic
medium is assumed to split into  two perfect fluid parts,
\begin{equation}
T ^{ik} = T ^{ik}_{_{\left(1 \right)}}
+ T ^{ik}_{_{\left(2 \right)}} {\mbox{ , }}
\label{1}
\end{equation}
with
\begin{equation}
T^{ik}_{_{\left(A \right)}} = \rho_{_{\left(A \right)}} u^{i}u^{k}
+ p_{_{\left(A \right)}} h^{ik}{\mbox{ ,}} \; \; \;
\mbox{\ \ }(A = 1, 2)\ ,
\label{2}
\end{equation}
where
$\rho_{_{\left(A \right)}}$ denotes the energy density and
$p_{_{\left(A \right)}}$ the equilibrium pressure of
species-$A$ particles.
For simplicity we assume both components to share the
same four-velocity $u^{i}$, normalized so that
$u^{i} u_{i} = - 1$. The tensor
$h^{ik}=g^{ik} + u^{i}u^{k}$ projects any vectorial quantity
on the hypersurface orthogonal to $u^{i}$.
The particle flow vector $N_{_{\left(A \right)}}^{i}$
of species $A$ is defined as
\begin{equation}
N_{_{\left(A \right)}}^{i} = n_{_{\left(A \right)}}u^{i}{\mbox{ , }}
\label{3}
\end{equation}
where $n _{_{\left(A \right)}}$ is the particle number density.
We are interested in situations where neither the particle
numbers nor the energy-momenta of the components are
separately conserved, i.e., 
conversion of particles and exchange of energy and momentum
between the components are  admitted.
The balance laws for the particle numbers are
\begin{equation}
N _{_{\left(A \right)} ;i}^{i}
= \dot{n}_{_{\left(A \right)}}
+ \Theta n_{_{\left(A \right)}}
= n _{_{\left(A \right)}} \Gamma _{_{\left(A \right)}}
{\mbox{ , }}
\label{4}
\end{equation}
where $\Theta \equiv u^{i}_{;i}$ is the fluid expansion
and $\Gamma _{_{\left(A \right)}}$ is the rate of change of the  
number of
particles of species $A$.
There is particle production for
$\Gamma _{_{\left(A \right)}} > 0$ and
particle decay for
$\Gamma _{_{\left(A \right)}} < 0$, respectively.
For $\Gamma _{_{\left(A \right)}} = 0$
we have separate particle number conservation (``detailed balance''). 

Interactions between the fluid components amount to
the mutual exchange of energy and momentum.
Consequently, there will be no local energy-momentum
conservation for the subsystems separately.
Only the energy-momentum tensor of the system as a
whole is conserved.

Denoting the loss and source terms in the separate
balances by $t ^{i}_{_{\left(A \right)}}$,
we write
\begin{equation}
T ^{ik}_{_{\left(A \right)} ;k}
= - t _{_{\left(A \right)}}^{i} {\mbox{ , }}
\label{5}
\end{equation}
implying
\begin{equation}
\dot{\rho}_{_{\left(A \right)}}
+ \Theta\left(\rho_{_{\left(A \right)}}
+ p_{_{\left(A \right)}}\right)
= u _{a} t _{_{\left(A \right)}}^{a}
{\mbox{ , }}
\label{6}
\end{equation}
and
\begin{equation}
\left(\rho _{_{\left(A \right)}}
+ p _{_{\left(A \right)}}\right) \dot{u}^{a}
+ p _{_{\left(A \right)} ,k}h ^{ak}
= - h ^{a}_{i}t ^{i}_{_{\left(A \right)}} {\mbox{ .}}
\label{7}
\end{equation}
All the considerations to follow will be independent
of the specific structure of the $t _{_{\left(A \right)}}^{i}$.
In other words, there are no limitations on the strength
or the structure of the interaction.

Each component is governed by a separate Gibbs equation:
\begin{equation}
T _{_{\left(A \right)}} \mbox{d} s _{_{\left(A \right)}}
= \mbox{d} \frac{\rho _{_{\left(A \right)}}}
{n _{_{\left(A \right)}}}
+ p _{_{\left(A \right)}}
\mbox{d} \frac{1}{n _{_{\left(A \right)}}} {\mbox{ , }}
\label{8}
\end{equation}
where $T _{_{\left(A \right)}}$ and  $s _{_{\left(A \right)}}$ are  
the temperature
and  entropy per particle of species $A$, respectively.
Using the balances (\ref{4}) and (\ref{6}), one finds for the time
behavior of the entropy per particle
\begin{equation}
n _{_{\left(A \right)}} T _{_{\left(A \right)}}
\dot{s}_{_{\left(A \right)}} = u _{a} t _{_{\left(A \right)}}^{a}
- \left(\rho _{_{\left(A \right)}}
+ p _{_{\left(A \right)}}\right) \Gamma _{_{\left(A \right)}} {\mbox{ .}}
\label{9}
\end{equation}
With nonvanishing source terms in the balances for
$n _{_{\left(A \right)}}$ and $\rho _{_{\left(A \right)}}$, the  
change in the
entropy per particle is different from zero in general.

The equations of state of the fluid components
are assumed to have the general form
\begin{equation}
p_{_{\left(A \right)}} = p_{_{\left(A \right)}}
\left(n_{_{\left(A \right)}}, T_{_{\left(A \right)}}\right) \ ,
\mbox{\ \ }
\rho_{_{\left(A \right)}} = \rho_{_{\left(A \right)}}
\left(n_{_{\left(A \right)}},
T_{_{\left(A \right)}}\right){\mbox{ , }}
\label{10}
\end{equation}
i.e., 
particle number densities $n_{_{\left(A \right)}}$ and
temperatures $T_{_{\left(A \right)}}$
are regarded as the  basic
thermodynamical variables.
The temperatures of the fluids are different in general.

Differentiating the last of the relations (\ref{10}), using the balances 
(\ref{4}) and (\ref{6}) as well as
the general relation
\begin{equation}
\frac{\partial \rho_{_{\left(A \right)}}}{\partial n_{_{\left(A  
\right)}}}
=
\frac{\rho_{_{\left(A \right)}} + p_{_{\left(A \right)}}}
{n_{_{\left(A \right)}}}
- \frac{T_{_{\left(A \right)}}}{n_{_{\left(A \right)}}}
\frac{\partial p_{_{\left(A \right)}}}
{\partial T_{_{\left(A \right)}}} {\mbox{ , }}
\label{11}
\end{equation}
that follows from the requirement that the entropy is a state
function,
we find the following expression for the temperature
behavior \cite{Calv}:
\begin{eqnarray}
\dot{T}_{_{\left(A \right)}}  &=& - T_{_{\left(A \right)}}
\left(\Theta - \Gamma _{_{\left(A \right)}} \right)
\frac{\partial p_{_{\left(A \right)}}}
{\partial \rho_{_{\left(A \right)}}}\nonumber\\
&& \mbox{\ \ }+ \frac{u _{a} t _{_{\left(A \right)}}^{a}
- \Gamma _{_{\left(A \right)}} \left(\rho _{_{\left(A \right)}}
+ p _{_{\left(A \right)}}\right)}
{\partial \rho _{_{\left(A \right)}}/ \partial T _{_{\left(A \right)}}}
{\mbox{ .}}
\label{12}
\end{eqnarray}
Here we have used the abbreviations
\[
\frac{\partial{p _{_{\left(A \right)}}}}
{\partial{\rho _{_{\left(A \right)}}}} \equiv
\frac{\partial p_{_{\left(A \right)}}
/\partial T_{_{\left(A \right)}}}{\partial \rho_{_{\left(A \right)}}/
\partial T_{_{\left(A \right)}}} \ ,
\mbox{\ \ }
\frac{\partial \rho_{_{\left(A \right)}}}{\partial n_{_{\left(A  
\right)}}}
\equiv   \left(\frac{\partial \rho_{_{\left(A \right)}}}{\partial  
n_{_{\left(A \right)}}} \right)_{T _{_{\left(A \right)}}}\ ,
\]
\[
\frac{\partial p_{_{\left(A \right)}}}
{\partial T_{_{\left(A \right)}}}
\equiv   \left(\frac{\partial p_{_{\left(A \right)}}}
{\partial T_{_{\left(A \right)}}} \right)_{n _{_{\left(A \right)}}} 
{\rm etc.}
\]
Both the source terms $\Gamma _{_{\left(A \right)}}$ and
$u _{a}t _{_{\left(A \right)}}^{a}$
in Eqs. (\ref{4}) and (\ref{6}) back react on the temperature.
The numerator of the second term on the right hand side of Eq.  
(\ref{12})
coincides with the right-hand side of Eq. (\ref{9}), i.e., the  
corresponding
terms disappear in the special case $\dot{s}_{_{\left(A \right)}} = 0$. 

For
$\Gamma _{_{\left(A \right)}} = u _{a}t _{_{\left(A \right)}}^{a} =  
0$ and
with $\Theta = 3\dot{a}/a$, where $a$ is a characteristic length  
scale (which in the homogeneous and isotropic case coincides with
the scale factor of the
Robertson-Walker metric), the equations of state 
$p _{_{\left(2 \right)}} = n _{_{\left(2 \right)}}T_{_{\left(2  
\right)}}$, $\rho_{_{\left(2 \right)}} = 3n_{_{\left(2  
\right)}}T_{_{\left(2 \right)}}$ for radiation
(subindex $\left(2 \right)$)
reproduce the well-known $T_{_{\left(2 \right)}} \sim a^{-1}$ behavior. 
With  $p_{_{\left(1 \right)}} = n_{_{\left(1 \right)}}T_{_{\left(1  
\right)}}$, $\rho_{_{\left(1 \right)}} = n_{_{\left(1 \right)}}m +
\frac{3}{2}n_{_{\left(1 \right)}}T_{_{\left(1 \right)}}$ one recovers 
$T_{_{\left(1 \right)}} \sim a^{-2}$ 
for matter (subindex $\left(1 \right)$). 

The entropy flow vector $S _{_{\left(A \right)}}^{a}$ is defined by 
\begin{equation}
S _{_{\left(A \right)}}^{a} = n _{_{\left(A \right)}}
s _{_{\left(A \right)}} u ^{a} {\mbox{ , }}
\label{13}
\end{equation}
and the contribution of component $A$ to the entropy
production density becomes
\begin{eqnarray}
S _{_{\left(A \right)} ;a}^{a} &=& n _{_{\left(A \right)}}
s _{_{\left(A \right)}} \Gamma _{_{\left(A \right)}} + n  
_{_{\left(A \right)}}
\dot{s}_{_{\left(A \right)}}\nonumber\\
&=& \left(s _{_{\left(A \right)}}
- \frac{\rho _{_{\left(A \right)}} + p _{_{\left(A \right)}}}
{n _{_{\left(A \right)}}T _{_{\left(A \right)}}}
\right)
n _{_{\left(A \right)}}\Gamma _{_{\left(A \right)}}
+ \frac{u _{a} t _{_{\left(A \right)}}^{a}}{T _{_{\left(A \right)}}} 
{\mbox{ , }}
\label{14}
\end{eqnarray}
where relation (\ref{9}) has been used.

According to the balances (\ref{5}) the condition of energy-momentum 
conservation for the system as a whole,
\begin{equation}
\left(T _{_{\left(1 \right)}}^{ik}
+ T _{_{\left(2 \right)}}^{ik}\right)_{;k} = 0 {\mbox{ , }}
\label{15}
\end{equation}
implies
\begin{equation}
t _{_{\left(1 \right)}}^{a} = - t _{_{\left(2 \right)}}^{a} {\mbox{ .}}
\label{16}
\end{equation}
There is no corresponding condition, however, for the
particle number balance as a whole.
Defining the integral particle number density $n$ as
$n = n _{_{\left(1 \right)}} + n _{_{\left(2 \right)}}$, we have  
\begin{equation}
\dot{n} + \Theta n = n \Gamma
 = n _{_{\left(1 \right)}}\Gamma _{_{\left(1 \right)}}
+ n _{_{\left(2 \right)}}\Gamma _{_{\left(2 \right)}}
{\mbox{ .}}
\label{17}
\end{equation}
$\Gamma$ is the rate by which the total particle number
$n$ changes.
We do {\it not} require $\Gamma$ to be zero since total
particle number conservation is only a very special case.

The entropy per particle is \cite{Groot}
\begin{equation}
s _{_{\left(A \right)}} = \frac{\rho _{_{\left(A \right)}}
+ p _{_{\left(A \right)}}}{n _{_{\left(A \right)}}T _{_{\left(A  
\right)}}}
- \frac{\mu _{_{\left(A \right)}}}
{T _{_{\left(A \right)}}}
{\mbox{ , }}
\label{18}
\end{equation}
where $\mu _{_{\left(A \right)}}$ is the chemical potential of  
species $A$.
Introducing the expression (\ref{18}) into Eq. (\ref{14}) yields
\begin{equation}
S ^{a}_{_{\left(A \right)} ;a} =
- \frac{\mu _{_{\left(A \right)}}}{T _{_{\left(A \right)}}}
n _{_{\left(A \right)}}\Gamma _{_{\left(A \right)}}
+ \frac{u _{a}t ^{a}_{_{\left(A \right)}}}{T _{_{\left(A \right)}}}
 {\mbox{ .}}
\label{19}
\end{equation}
For the total entropy production density
$S ^{a}_{;a} = S ^{a}_{_{\left(1 \right)} ;a}
+ S ^{a}_{_{\left(2 \right)} ;a}$
we obtain
\begin{eqnarray}
S ^{a}_{;a} &=& - \frac{\mu _{_{\left(2 \right)}}}
{T _{_{\left(2 \right)}}}n \Gamma
- \left(\frac{\mu _{_{\left(1 \right)}}}
{T _{_{\left(1 \right)}}}
- \frac{\mu _{_{\left(2 \right)}}}{T _{_{\left(2 \right)}}}
\right)
n _{_{\left(1 \right)}}\Gamma _{_{\left(1 \right)}} \nonumber\\
&& \mbox{\ \ \ \ \ \ \ \ \ \ }+ \left(\frac{1}{T _{_{\left(1 \right)}}} 
- \frac{1}{T _{_{\left(2 \right)}}}\right)u _{a}
t _{_{\left(1 \right)}}^{a} {\mbox{ .}}
\label{20}
\end{eqnarray}
The condition $S ^{a}_{;a} = 0$ requires the well-known
equilibrium
conditions (see, e.g., \cite{Bern} (chapter 5))
\begin{equation}
\mu _{_{\left(1 \right)}} = \mu _{_{\left(2 \right)}}   {\mbox{ , }}   
T _{_{\left(1 \right)}} = T _{_{\left(2 \right)}} {\mbox{ , }}
\label{21}
\end{equation}
as well as $\Gamma = 0$.

\section{Basic relations for an evaporating black hole component}
In this section we apply the formalism outlined above to a  
mixture
of massless radiation (component 2) and evaporating PBHs
(component 1). We assume all the black holes of
the same mass $m_{_{\left(BH \right)}}$, and
that they evaporate solely into
radiation particles, though this is admittedly  a rough
approximation \cite{Page1}.
The PBH component is treated as a nonrelativistic fluid with
\begin{equation}
p _{_{\left(BH \right)}} = 0
\label{22}
\end{equation}
and the energy density
\begin{equation}
\rho _{_{\left(BH \right)}} = n _{_{\left(BH \right)}}
m _{_{\left(BH \right)}}\ .
\label{23}
\end{equation}
The number $N _{_{\left(BH \right)}}$ of PBHs in a comoving volume  
$a ^{3}$,
$N _{_{\left(BH \right)}} = n _{_{\left(BH \right)}}a ^{3}$, is not  
preserved and, according to Eq. (\ref{4}), we may write down a  
balance equation for the corresponding PBH number flow vector
$N _{_{\left(BH \right)}}^{i} = n _{_{\left(BH \right)}}u ^{i}$,
\begin{equation}
N _{_{\left(BH \right)} ;i}^{i} =
\dot{n}_{_{\left(BH \right)}} + \Theta n_{_{\left(BH \right)}} =
n _{_{\left(BH \right)}} \Gamma _{_{\left(BH \right)}}
{\mbox{ . }}
\label{24}
\end{equation}
>From Eq. (\ref{23})
the black hole energy balance becomes [cf. Eq. (\ref{6})]
\begin{equation}
\dot{\rho }_{_{\left(BH \right)}} + \Theta \rho _{_{\left(BH \right)}} = 
u _{a}t ^{a}_{_{\left(BH \right)}}
\label{25}
\end{equation}
with
\begin{equation}
u _{a}t ^{a}_{_{\left(BH \right)}} = \rho _{_{\left(BH \right)}}
\left[\Gamma _{_{\left(BH \right)}}
+ \frac{\dot{m}_{_{\left(BH \right)}}}
{m _{_{\left(BH \right)}}}\right] \ .
\label{26}
\end{equation}
The entropy per black hole is
\begin{equation}
s _{_{\left(BH \right)}} = 4 \pi  m _{_{\left(BH \right)}}^{2}
{\mbox{,}}
\label{27}
\end{equation}
while the black hole temperature is related to its mass by
the well-known formula \cite{HAW2}
\begin{equation}
T _{_{\left(BH \right)}} = \frac{1}{8 \pi m _{_{\left(BH \right)}}}
\ .
\label{28}
\end{equation}
The temperature $T _{_{\left(BH \right)}}$ is attributed to each  
PBH individually, i.e., it is {\it not} a conventional fluid  
temperature.
With this basic conceptional difference in mind it is nevertheless  
possible to associate  standard fluid-type quantities to the  
single-mass PBH component.
Specifying the general relation (\ref{18}) to the case at hand,
a chemical potential may be associated with
the black hole component by
\begin{equation}
\mu _{_{\left(BH \right)}} = \frac{\rho _{_{\left(BH \right)}}}
{n _{_{\left(BH \right)}}} - T _{_{\left(BH \right)}}
s _{_{\left(BH \right)}}
= \frac{1}{2} m _{_{\left(BH \right)}}\ .
\label{29}
\end{equation}
Together with relation (\ref{28}) the expression (\ref{23}) for the  
PBH energy density fits into the general structure given in Eqs.  
(\ref{10}).
The corresponding partial derivatives are
\begin{equation}
\frac{\partial{\rho _{_{\left(BH \right)}}}}
{\partial{T _{_{\left(BH \right)}}}}
= - \frac{\rho _{_{\left(BH \right)}}}{T _{_{\left(BH \right)}}}\ ,  
\ \ \
\frac{\partial{\rho _{_{\left(BH \right)}}}}
{\partial{n _{_{\left(BH \right)}}}}
= m_{_{\left(BH \right)}}\ .
\label{30}
\end{equation}
Furthermore, the relationship
\begin{equation}
\frac{\partial{\mu _{_{\left(BH \right)}}}}
{\partial{T _{_{\left(BH \right)}}}} =
- \frac{1}{2} \frac{m _{_{\left(BH \right)}}}{T _{_{\left(BH  
\right)}}}\ ,
\label{31}
\end{equation}
holds.
It is essential that
the general temperature law (\ref{12}) is applicable as well and  
specifies
[cf. Eqs. (\ref{22}) and (\ref{26})] to
\begin{equation}
\dot{T}_{_{\left(BH \right)}}  =
\frac{u _{a} t _{_{\left(BH \right)}}^{a}
- \Gamma _{_{\left(BH \right)}} \rho _{_{\left(BH \right)}}}
{\partial \rho _{_{\left(BH \right)}}/ \partial T _{_{\left(BH  
\right)}}}
= - T _{_{\left(BH \right)}} \frac{\dot{m}_{_{\left(BH \right)}}}
{m _{_{\left(BH \right)}}}
{\mbox{ ,}}
\label{32}
\end{equation}
which is consistent with formula (\ref{28}).
{\it The BH temperature behavior
$T _{_{\left(BH \right)}} \propto m _{_{\left(BH \right)}}^{-1}$  
may be regarded  as a special case of the general fluid temperature  
law (\ref{12}).}
This provides the main motivation for our fluid approach to PBHs. 
On the other hand, this approach relies on the possibility of treating 
the black holes as an ensemble of noninteracting particles. 
The validity of the latter assumption can be shown by a simple
Newtonian argument \cite{GD}. 
The relative velocity of neighboring black holes due to the Hubble
expansion is 
\[
v_{exp} = \frac{\mbox{d}}{\mbox{d}t}
\left(n _{_{\left(BH \right)}}^{-1/3}\right)\ .
\]
With the help of Eq. (\ref{24}) we obtain 
\[
v_{exp} = \frac{\Theta}{3}n _{_{\left(BH \right)}}^{-1/3}
\left(1 - \frac{\Gamma _{_{\left(BH \right)}}}{\Theta}\right)
\]
for this quantity. 
Since for a spatially flat FLRW universe
\[
\frac{\Theta}{3} = \sqrt{\frac{8\pi}{3}\left(n _{_{\left(BH \right)}} 
m _{_{\left(BH \right)}} 
+ \rho _{_{\left(2 \right)}}\right)}
\]
is valid, we find 
\[
v_{exp} = n _{_{\left(BH \right)}}^{-1/3}\sqrt{\frac{8\pi}{3}
\left(n _{_{\left(BH \right)}}
m _{_{\left(BH \right)}} + \rho _{_{\left(2 \right)}}\right)}
\left(1 - \frac{\Gamma _{_{\left(BH \right)}}}{\Theta} \right) \ . 
\]
The velocity required for a BH to escape the (Newtonian)
gravitational pull of its nearest neighbor is 
\[
v_{esc} = \sqrt{2m _{_{\left(BH \right)}}
n _{_{\left(BH \right)}}^{-1/3}} 
= n _{_{\left(BH \right)}}^{-1/3}
\sqrt{2n _{_{\left(BH \right)}}m _{_{\left(BH \right)}}}\ .
\]
Comparing the expressions for $v_{exp}$ and $v_{esc}$, we find that 
$v_{exp} > v_{esc}$, provided 
\[
\sqrt{\frac{4\pi}{3}\left(1 + 
\frac{\rho _{_{\left(2 \right)}}}
{n _{_{\left(BH \right)}}m _{_{\left(BH \right)}}}\right)}
\left(1 - \frac{\Gamma _{_{\left(BH \right)}}}{\Theta}\right) > 1\ .
\]
The latter relation is always satisfied for 
$\Gamma _{_{\left(BH \right)}} \leq 0$, the case of main interest here, but also for a
certain range of values $\Gamma _{_{\left(BH \right)}} > 0$. 
We conclude that the behavior of the individual BHs is primarily
determined by the expansion of the universe and not by their mutual
interaction. 
This justifies their treatment as a noninteracting particle species.

Let us assume from now on  that
the source terms on the right-hand side
of Eq. (\ref{9}) cancel among themselves, i.e., that the entropy
per particle of component 2 is preserved.
This isentropy condition
amounts to the assumption that the fluid-2 particles at any stage are 
amenable to a perfect fluid description.
With $\dot{s}_{_{\left(2 \right)}} = 0$, via Eq. (\ref{9}) equivalent to
\begin{equation}
u _{a} t ^{a}_{_{\left(2 \right)}} =
\left(\rho _{_{\left(2 \right)}} + p _{_{\left(2 \right)}}\right)
\Gamma _{_{\left(2 \right)}}
{\mbox{ , }}
\label{33}
\end{equation}
the expression (\ref{12}) for the temperature behavior simplifies   
considerably:
\begin{equation}
\dot{s}_{_{\left(2 \right)}} = 0
\quad \Rightarrow \quad
\frac{\dot{T}_{_{\left(2\right)}}}{T_{_{\left(2\right)}}}  = -
\left(\Theta - \Gamma _{_{\left(2 \right)}} \right)
\frac{\partial p_{_{\left(2 \right)}}}
{\partial \rho_{_{\left(2 \right)}}}\ .
\label{34}
\end{equation}

Combining Eqs. (\ref{16}) and (\ref{33}), one has
\begin{equation}
u _{a} t ^{a}_{_{\left(BH \right)}} = - u _{a} t ^{a}_{_{\left(2  
\right)}}
= - \left(\rho _{_{\left(2 \right)}} + p _{_{\left(2 \right)}}\right) 
\Gamma _{_{\left(2 \right)}}
{\mbox{ , }}
\label{35}
\end{equation}
and we find the expression
\begin{equation}
\Gamma _{_{\left(2 \right)}} = - \frac{\rho _{_{\left(BH \right)}}}
{\rho _{_{\left(2 \right)}} + p _{_{\left(2 \right)}}}
\left[\Gamma _{_{\left(BH \right)}}
+ \frac{\dot{m}_{_{\left(BH \right)}}}
{m _{_{\left(BH \right)}}}\right]
\label{36}
\end{equation}
for the rate of change of the fluid-2 particle number.
Negative rates $\Gamma _{_{\left(BH \right)}}$ and
$\dot{m}_{_{\left(BH \right)}}/m _{_{\left(BH \right)}}$  
(evaporation) imply a positive $\Gamma _{_{\left(2 \right)}}$ and  
vice versa.

The contribution of the black hole component to the entropy
production density is [cf. Eqs. (\ref{13}) and (\ref{24})]
\begin{equation}
S _{_{\left(BH \right)}\ ;a}^{a} = n _{_{\left(BH \right)}}
s _{_{\left(BH \right)}}
\left[\Gamma _{_{\left(BH \right)}}
+ \frac{\dot{s}_{_{\left(BH \right)}}}{s _{_{\left(BH \right)}}}\right]
\label{37}
\end{equation}
with
\begin{equation}
\frac{\dot{s}_{_{\left(BH \right)}}}{s _{_{\left(BH \right)}}} =
2 \frac{\dot{m}_{_{\left(BH \right)}}}{m _{_{\left(BH \right)}}}\ ,
\label{38}
\end{equation}
according to Eq. (\ref{27}).
The fluid-2 component contributes with
\begin{equation}
S _{_{\left(2 \right)}\ ;a}^{a}
= n _{_{\left(2 \right)}} s _{_{\left(2 \right)}}
\Gamma _{_{\left(2 \right)}}\ .
\label{39}
\end{equation}
Assuming now the second component to be radiation with
$p _{_{\left(2 \right)}} = \rho _{_{\left(2 \right)}}/3$ and
$\mu _{_{\left(2 \right)}} = 0$, the overall entropy production  
density (\ref{20}) becomes
\begin{eqnarray}
S ^{a}_{; a} &=& \rho _{_{\left(BH \right)}} \Gamma _{_{\left(BH  
\right)}}
\left[\frac{1}{2 T _{_{\left(BH \right)}}}
- \frac{1}{T _{_{\left(2 \right)}}}\right] \nonumber\\
&&+ \rho _{_{\left(BH \right)}} \frac{\dot{m}_{_{\left(BH \right)}}}
{m _{_{\left(BH \right)}}}
\left[\frac{1}{T _{_{\left(BH \right)}}}
- \frac{1}{T _{_{\left(2 \right)}}}\right]\ ,
\label{40}
\end{eqnarray}
where we have used relations (\ref{39}) and (\ref{36}).
This formula for the entropy production density is still completely  
general and, within the fluid picture, holds for PBH formation
($\Gamma _{_{\left(BH \right)}} > 0$) and mass accretion
($\dot{m}_{_{\left(BH \right)}} > 0$) as well as for the  
evaporation process
($\Gamma _{_{\left(BH \right)}} < 0$ and
$\dot{m}_{_{\left(BH \right)}} < 0$).
For $T _{_{\left(2 \right)}} > 2 T _{_{\left(BH \right)}}$ the  
second law favors the formation of PBHs ($\Gamma _{_{\left(BH  
\right)}} > 0$) and mass accretion
($\dot{m}_{_{\left(BH \right)}} > 0$), while for
$T _{_{\left(2 \right)}} <  T _{_{\left(BH \right)}}$ the reverse  
processes dominate.

Let us assume that at some initial time  $t _{0}$ the black hole  
temperature coincides with the radiation temperature, i.e.,
$T _{_{\left(BH \right)}}\left(t _{0} \right)
= T _{_{\left(2 \right)}}\left(t _{0} \right)$.
Under this condition the second term in Eq. (\ref{40}) vanishes and
the entropy production density 
reduces to
\begin{equation}
S ^{a}_{; a} = - \frac{1}{2}
\frac{\rho _{_{\left(BH \right)}}}{T _{_{\left(BH \right)}}}
\Gamma _{_{\left(BH \right)}}
= - n _{_{\left(BH \right)}} s _{_{\left(BH \right)}}
\Gamma _{_{\left(BH \right)}}
\label{41}
\end{equation}
at $t = t _{0}$.
To obtain the second relation (\ref{41}) we have used Eqs.
(\ref{27}) and (\ref{28}).
Obviously, this case requires $\Gamma _{_{\left(BH \right)}} < 0$  
in order to satisfy $S ^{a}_{;a} > 0$, i.e., decay
($\Gamma _{_{\left(BH \right)}} < 0$) of the PBH component.
In other words,
{\it the formation of PBHs ($\Gamma _{_{\left(BH \right)}} > 0$) is  
thermodynamically forbidden if the temperature of the PBHs  
coincides with the temperature of the ambient radiation, i.e., for
$T _{_{\left(BH \right)}} = T _{_{\left(2 \right)}}$.} 

Let us now consider the second term in Eq. (\ref{40}). 
This contribution, which vanishes for 
$T _{_{\left(BH \right)}}\left(t_{0}\right) 
= T _{_{\left(2 \right)}}\left(t_{0}\right)$, represents the
entropy production density in case the number of PBHs is constant,
i.e., for $\Gamma _{_{\left(BH \right)}} = 0$. 
Restricting ourselves momentarily to a fixed BH number, there are
two possibilities for the subsequent evolution of the system. 
For $T _{_{\left(2 \right)}} < T _{_{\left(BH \right)}}$ 
immediately after $t = t_{0}$ a positive entropy
production requires 
$\dot{m}_{_{\left(BH \right)}}/m _{_{\left(BH \right)}} < 0$, i.e., PBH evaporation. 
This process will be studied in some detail below. 
For $T _{_{\left(BH \right)}} < T _{_{\left(2 \right)}}$ on the other 
hand, the condition 
$S^{a}_{;a} > 0$ is only satisfied for 
$\dot{m}_{_{\left(BH \right)}}/m _{_{\left(BH \right)}} > 0$, i.e., for mass accretion. 
Becoming larger, the BHs also become colder [cf. Eq. (\ref{28})] and
their growth seems to continue indefinitely, i.e., until they have
eaten up the entire universe. 
However, such a kind of scenario is thermodynamically impossible as can
be seen by the following argument. 
The essential point is that the BH mass accretion back reacts on the
temperature of the ambient radiation. 
For an accretion rate 
$\dot{m}_{_{\left(BH \right)}}/m _{_{\left(BH \right)}} > 0$ the rate (\ref{36}) is negative (recall
that 
$\Gamma _{_{\left(BH \right)}} = 0$ in the present discussion). 
The cooling rate (\ref{34}) of the radiation temperature then becomes
\[
\frac{\dot{T}_{_{\left(2 \right)}}}{T _{_{\left(2 \right)}}} = - \frac{1}{3}
\left(\Theta + \frac{3}{4}\frac{\rho _{_{\left(BH \right)}}}
{\rho _{_{\left(2 \right)}}}
\frac{\dot{m}_{_{\left(BH \right)}}}{m _{_{\left(BH \right)}}}
\right) \ .
\]
It is obvious that from some time on the temperature 
$T _{_{\left(2 \right)}}$ will
cool off faster than $T _{_{\left(BH \right)}}$ [cf. Eq. (\ref{32})]. 
Consequently, $T _{_{\left(2 \right)}}$ will approach 
$T _{_{\left(BH \right)}}$. 
As soon as $T _{_{\left(2 \right)}}$ has fallen below 
$T _{_{\left(BH \right)}}$, mass accretion stops
since for $T _{_{\left(2 \right)}} < T _{_{\left(BH \right)}}$ the rate 
$\dot{m}_{_{\left(BH \right)}}/m _{_{\left(BH \right)}}$ has to be negative in order to guarantee a
positive entropy production, i.e., the process now proceeds in the
reverse direction. 
We conclude that the second law of thermodynamics forbids a
catastrophic growth of the PBHs. 
These considerations may also be regarded as a justification of the 
``initial" condition 
$T _{_{\left(BH \right)}}\left(t_{0}\right) = 
T _{_{\left(2 \right)}}\left(t_{0}\right)$ for the
evaporation process. 

Now we continue discussing the case with $\Gamma_{BH}$ different from
zero. 
The situation $S ^{a}_{;a} = 0$ generally corresponds to a ``detailed balance'' at 
$t = t _{0}$, i.e. both reactions, the formation of PBHs and their  
evaporation, proceed at the same rate such that there is no change  
in the net numbers $N _{_{\left(BH \right)}}$ and
$N _{_{\left(2 \right)}}$.
This is equivalent to
$\Gamma _{_{\left(BH \right)}}\left(t _{0} \right) =
\dot{m} _{_{\left(BH \right)}}\left(t _{0} \right)
/m _{_{\left(BH \right)}}\left(t _{0} \right)
= \Gamma _{_{\left(2 \right)}}\left(t _{0} \right) = 0$.

For $t < t _{0}$ the formation of PBHs and accretion, i.e.,
$\Gamma _{_{\left(BH \right)}} >  0$,
$\dot{m} _{_{\left(BH \right)}}/m _{_{\left(BH \right)}} >  0$
and $\Gamma _{_{\left(2 \right)}} < 0$ are thermodynamically favored. 
For $t > t _{0}$ the PBH evaporation with
$\Gamma _{_{\left(BH \right)}} <  0$,
$\dot{m} _{_{\left(BH \right)}}/m _{_{\left(BH \right)}} <  0$
and $\Gamma _{_{\left(2 \right)}} > 0$ dominates.

We are interested here in the second part of this process, i.e., in  
the evolution of the universe for $t \geq t _{0}$.
We will not discuss issues of PBH formation (see the introduction  
for possible mechanisms)
but simply assume that at some early time a considerable amount of  
cosmic matter was in the form of PBHs of the same mass and that this  
component subsequently (e.g., until the epoch of nuleosynthesis)  
decayed.
We will investigate the thermodynamic aspects of this evaporation  
process and its implications for the cosmological dynamics.

According to Eq. (\ref{36}) negative values of the rates
$\Gamma _{_{\left(BH \right)}}$ and
$\dot{m} _{_{\left(BH \right)}}/m _{_{\left(BH \right)}}$ imply a  
positive
$\Gamma _{_{\left(2 \right)}}$.
The production rate $\Gamma _{_{\left(2 \right)}}$ may either be  
larger or smaller than the expansion rate $\Theta $.
For $\Gamma _{_{\left(2 \right)}} < \Theta $ the fluid temperature  
decreases according to Eq. (\ref{34}), while the BH temperature  
increases according to Eq. (\ref{32}).
It follows that $T _{_{\left(BH \right)}} > T _{_{\left(2 \right)}}$ 
at $t > t _{0}$. The evaporation process will continue since
$T _{_{\left(2 \right)}} < T _{_{\left(BH \right)}}$ requires
$\Gamma _{_{\left(BH \right)}} < 0$ and
$\dot{m}_{_{\left(BH \right)}}/m _{_{\left(BH \right)}} < 0$ to  
guarantee
$S ^{a}_{;a} > 0$ in Eq. (\ref{40}).

For $\Gamma _{_{\left(2 \right)}} >
\Theta $, however, hypothetically realized e.g. by a large initial ratio 
$\rho _{_{\left(BH \right)}}/ \rho _{_{\left(2 \right)}}$,
the fluid temperature increases.
If this increase is smaller than the increase in $T _{_{\left(BH  
\right)}}$
we have again
$T _{_{\left(2 \right)}} < T _{_{\left(BH \right)}}$ and the PBH  
evaporation goes on since it remains thermodynamically favorable
($S ^{a}_{;a} > 0$).
But an increase in $T _{_{\left(2 \right)}}$ stronger than that in
$T _{_{\left(BH \right)}}$ results in a fluid temperature which is  
higher than $T _{_{\left(BH \right)}}$.
For $T _{_{\left(2 \right)}} > 2 T _{_{\left(BH \right)}}$ a  
positive entropy production (\ref{40}) requires
$\Gamma _{_{\left(BH \right)}} > 0$ and
$\dot{m}_{_{\left(BH \right)}}/m _{_{\left(BH \right)}} > 0$,  
implying a quick transition to a negative $\Gamma _{_{\left(2  
\right)}}$, i.e., the process confines itself.
A strong ``reheating'' of the fluid will stop the evaporation and  
reverse the process.
Now, the second law requires PBHs to be formed out of the radiation  
and accrete mass.
A negative $\Gamma _{_{\left(2 \right)}}$, on the other hand, will make 
$T _{_{\left(2 \right)}}$ subsequently decrease [cf. Eq. (\ref{34})]. 
If $T _{_{\left(2 \right)}}$ has fallen below $T _{_{\left(BH  
\right)}}$, the evaporation process may continue 
(see the discussion following Eq. (\ref{41})).

We conclude that {\it the PBH evaporation is a self-confining  
process and, consequently,
the accompanying ``reheating'' of the radiation is limited on  
general thermodynamical grounds.}

To study the corresponding dynamics of the two-component system in  
more detail
we resort to the  decay law \cite{HAW2}
\begin{equation}
\dot{m}_{_{\left(BH \right)}} = - \frac{A}{m _{_{\left(BH  
\right)}}^{2}}\ ,
\label{42}
\end{equation}
where $A$ may be taken as a positive-definite constant
to be determined by the condition
$m _{_{\left(BH \right)}}\left(t _{0} + \tau  \right) = 0$,
with $\tau $ the lifetime of a Schwarzschild black hole that
obeys a law of the type (\ref{42}) from its formation up to
its final dissapearance, leaving no remnant behind.
We assume that all PBHs start evaporating at the same time and that 
accretion of the ambience
matter on the black holes can be ignored \cite{CHW,GD}.
In reality
Eq. (\ref{42}) is an approximation to Hawking's law since $A$ is
not rigorously constant over the entire radiation period, but
depends on the number of particle species emitted
by the black hole, whence it slowly increases with the
inverse of the black hole mass (see \cite{Page1} and
\cite{TZP}).
For the rate
$\dot{m}_{_{\left(BH \right)}}/m _{_{\left(BH \right)}}$
one finds
\begin{equation}
\frac{\dot{m}_{_{\left(BH \right)}}}{m _{_{\left(BH \right)}}}
= - \frac{1}{3 \tau }\left(\frac{m _{_{\left(BH \right)}}
\left(t _{0} \right)}
{m _{_{\left(BH \right)}}\left(t \right)} \right)^{3}\ .
\label{43}
\end{equation}
Integration of this equation yields
\begin{equation}
m _{_{\left(BH \right)}}\left(t \right) =
m _{_{\left(BH \right)}}\left(t _{0} \right)
\left[1 - \frac{t - t _{0}}{\tau } \right]^{1/3}\ .
\label{44}
\end{equation}
The last expression allows one to write the evaporation rate  
(\ref{43}) as
\begin{equation}
\frac{\dot{m}_{_{\left(BH \right)}}}{m _{_{\left(BH \right)}}}
= - \frac{1}{3 \tau }\frac{1}{1 - \frac{t - t _{0}}{\tau }}\ .
\label{45}
\end{equation}
The number decay rate of the PBHs is determined by their inverse  
lifetime, i.e.,
\begin{equation}
\Gamma _{_{\left(BH \right)}} = - \frac{1}{\tau }\ .
\label{46}
\end{equation}
Formulas (\ref{45}) and (\ref{46}) hold if the PBHs are purely  
decaying with the reverse process entirely suppressed.

In order to describe the process of formation and subsequent  
evaporation of PBHs in full detail, one needs explicit expressions  
for the rates $\Gamma _{_{\left(BH \right)}}$ and
$\dot{m}_{_{\left(BH \right)}}/m _{_{\left(BH \right)}}$ through  
the transition period from predominant PBH formation to the  
evaporation phase, where
$\Gamma _{_{\left(BH \right)}}$ and
$\dot{m}_{_{\left(BH \right)}}/m _{_{\left(BH \right)}}$ switch  
from positive to negative values.
While our general framework up to Eq. (\ref{40}) is able to cover  
this most general case, the simple expressions (\ref{45}) and  
(\ref{46}) for
$\dot{m}_{_{\left(BH \right)}}/m _{_{\left(BH \right)}}$ and
$\Gamma _{_{\left(BH \right)}}$, respectively, are valid only for  
pure evaporation.

To simplify our description we will assume the interval between the  
time at which $\Gamma _{_{\left(BH \right)}}$ changes its sign and  
the time at which the expressions (\ref{45}) and (\ref{46}) are  
valid to be
negligibly small, so that
it may be justified to approximately identify both times.
Under this condition the simple expressions (\ref{45}) and (\ref{46}) 
may be used for the whole range $t \geq t _{0}$.
As a consequence, we have to deal with a nonzero decay rate already at 
$t = t _{0}$ and the ``initial'' entropy production rate is  
positive, according to Eq. (\ref{41}).

Applying the simple rates (\ref{45}) and (\ref{46}) from $t _{0}$  
on will provide us with a transparent picture of the evaporation  
process, although it implies the assumption of an equality of $T  
_{_{\left(BH \right)}}$ and
$T _{_{\left(2 \right)}}$ at a time $t _{0}$ somewhat later than  
the ``real'' beginning
($\Gamma _{_{\left(BH \right)}} =
\dot{m}_{_{\left(BH \right)}}/m _{_{\left(BH \right)}} = 0$)
of the evaporation process.

Assuming $p _{_{\left(2 \right)}} = \rho _{_{\left(2 \right)}}/3$  
in Eq. (\ref{36}), the production rate $\Gamma _{_{\left(2  
\right)}}$ at $t = t _{0}$ is
\begin{equation}
\Gamma _{_{\left(2 \right)}}\left(t _{0} \right)
= \frac{\beta }{\tau }\ ,
\mbox{\ \ \ \ \ \ \ }
\beta \equiv  \frac{\rho _{_{\left(BH \right)}}\left(t _{0} \right)}
{\rho _{_{\left(2 \right)}}\left(t _{0} \right)}\ ,
\label{47}
\end{equation}
where $\beta $ is the initial ratio of the PBH energy density to  
the radiation energy density.
Given the PBH lifetime $\tau $, the initial production rate is  
determined by the ratio of the energy densities.
According to Eq. (\ref{34}) this rate fixes the subsequent  
behaviour of $T _{_{\left(2 \right)}}$.
It is obvious from Eq. (\ref{34}), that for a rate
$\Gamma _{_{\left(2 \right)}} = 3 H
- 3 \dot{m}_{_{\left(BH \right)}}/m _{_{\left(BH \right)}}$, where
$H \equiv  \frac{\Theta }{3}$ is the Hubble parameter,
the fluid temperature $T _{_{\left(2 \right)}}$ increases at the  
same rate as
$T _{_{\left(BH \right)}}$ [cf. Eq. (\ref{32})].
Since evaporation is only possible for
$T _{_{\left(2 \right)}} < T _{_{\left(BH \right)}}$, this is the  
highest rate for which there is evaporation.
A still higher rate interrupts the evaporation and  
thermodynamically favors a temporary PBH formation phase until $T  
_{_{\left(2 \right)}} \approx
T _{_{\left(BH \right)}}$ again (see the discussion following Eq.  
(\ref{41})).
Combining this maximum rate at $t = t _{0}$ with the rate (\ref{45}) at 
$t = t _{0}$, provides us with the condition for evaporation,
\begin{equation}
\beta \leq 3 H _{0}\tau + 1 \ ,
\label{49}
\end{equation}
where $H _{0}$ is the Hubble parameter at $t _{0}$.
This condition limits the initial ratio $\beta $.
For $\tau \leq H _{0}^{-1}$ the PBHs evaporate within one Hubble  
time and $\beta $ is restricted to $\beta \leq 4$.
If the BH lifetime is much larger than the Hubble time, i.e.,
$\tau \gg H _{0}^{-1}$, the initial PBH abundance may be larger, i.e. 
$\rho _{_{\left(BH \right)}} \gg \rho _{_{\left(2 \right)}}$  
corresponding to
$\Gamma _{_{\left(2 \right)}} \gg \Gamma _{_{\left(BH \right)}}$,
is possible at $t = t _{0}$.

Notice that these are restrictions following from the  
thermodynamics of the decay process.
We did not take into account here any limits on the fractional  
abundance from PBH formation mechanisms which might be more  
restrictive than the present ones.

We may define a ``reheating'' temperature
$T _{_{\left(2 \right)}}^{^{reh}}$ as the maximum temperature for  
the radiation component by
$\dot{T}_{_{\left(2 \right)}}^{^{reh}} = 0$.
According to Eq. (\ref{34}) the corresponding condition is
$\Gamma _{_{\left(2 \right)}} = \Theta $.
With $\rho _{_{\left(2 \right)}} = 3 n _{_{\left(2 \right)}}
T _{_{\left(2 \right)}}$ and Eq. (\ref{47}), i.e., assuming the  
reheating to proceed in the initial phase of the evaporation  
process, we find
\begin{equation}
T _{_{\left(2 \right)}}^{^{reh}} \approx \frac{1}{3}
\frac{n _{_{\left(BH \right)}}}{n _{_{\left(2 \right)}}}
\frac{m _{_{\left(BH \right)}}}{\Theta \tau }\ .
\label{50}
\end{equation}

The energy balance (\ref{25}) with the ``sink'' (\ref{26}), applied  
to the evaporation process,  becomes
\begin{equation}
\dot{\rho} _{_{\left(BH \right)}} + \Theta \rho _{_{\left(BH \right)}} 
=  \rho _{_{\left(BH \right)}}
\left[\Gamma _{_{\left(BH \right)}} +
\frac{\dot{m}_{_{\left(BH \right)}}}{m _{_{\left(BH \right)}}} \right]
\label{51}
\end{equation}
with the rates (\ref{46}) and (\ref{45}).
For the radiation component we have
\begin{equation}
\dot{\rho} _{_{\left(2 \right)}} + \frac{4}{3}\Theta
\rho _{_{\left(2 \right)}}
= - \rho _{_{\left(BH \right)}}
\left[\Gamma _{_{\left(BH \right)}} +
\frac{\dot{m}_{_{\left(BH \right)}}}{m _{_{\left(BH \right)}}} \right]\ .
\label{52}
\end{equation}

With respect to ``equilibrium'' solutions found by Barrow {\em et  
al.}  \cite{BCL}, in which the ratio of the energy density in the  
PBH component to the energy density of the ambient radiation remains  
constant, we address the question whether a corresponding behavior  
is possible in the present context.
To this purpose we have to study whether a relation
$\rho _{_{\left(BH \right)}} = \tilde{\beta }
\rho _{_{\left(2 \right)}}$ with $\tilde{\beta }$ = const is  
compatible with the equations (\ref{51}) and (\ref{52}).
Introducing $\rho _{_{\left(BH \right)}} = \tilde{\beta }
\rho _{_{\left(2 \right)}}$ in either Eq. (\ref{51}) or Eq.  
(\ref{52}) provides us with a relation
\begin{equation}
\Gamma _{_{\left(2 \right)}} = \frac{1}{4}
\frac{\tilde{\beta }}{\tilde{\beta } + 1}\Theta
\label{53}
\end{equation}
with $\Gamma _{_{\left(2 \right)}} = - \frac{3}{4}
\tilde{\beta }\left[\Gamma _{_{\left(BH \right)}}
+ \dot{m}_{_{\left(BH \right)}}/m _{_{\left(BH \right)}} \right]$
[cf. Eq. (\ref{36})].
It follows that such kind of equilibrium solutions requires the rate 
$\Gamma _{_{\left(2 \right)}}$ (and, equivalently, the sum
$\Gamma _{_{\left(BH \right)}}
+ \dot{m}_{_{\left(BH \right)}}/m _{_{\left(BH \right)}}$) to be  
proportional to the expansion rate $\Theta  $.
But this is clearly incompatible with the expressions (\ref{45})
and (\ref{46}).
We conclude that equilibrium solutions do not exist in our  
two-fluid model.
This is at variance with Barrow {\em et al.} (see equation (24)
of Ref. \cite{BCL}) who (for a different PBH configuration) found a  
corresponding equilibrium behavior for the
evolution of $\rho_{_{\left( BH \right)}}$ and
$ \rho_{_{\left(2 \right)}}$.

Integration of equations (\ref{51}) and (\ref{52}) with the rates  
(\ref{45}) and (\ref{46}) provides us with the expressions
\begin{equation}
\rho _{_{\left(BH \right)}}\left(t \right)
= \rho _{_{\left(BH \right)}}\left(t _{0} \right)
\frac{a _{0}^{3}}{a ^{3}}
\left[1 - \frac{t - t _{0}}{\tau } \right]^{1/3}
\exp{\left[- \frac{t - t _{0}}{\tau } \right]}\ ,
\label{54}
\end{equation}
for the energy density of the black hole component,
and
\begin{eqnarray}
\rho _{_{\left(2 \right)}}\left(t \right)
&=& \rho _{_{\left(2 \right)}}\left(t _{0} \right)\frac{a  
_{0}^{4}}{a ^{4}}
+ \frac{1}{a ^{4}}
\frac{\rho _{_{\left(BH \right)}}
\left(t _{0} \right)a_{0}^{3}}{\tau}\times\nonumber\\
&&\mbox{\ \ }
\times\int_{t _{0}}^{t} \mbox{d}t\ a \left(t \right)
\frac{\frac{4}{3} - \frac{t - t _{0}}{\tau }}
{\left[1 - \frac{t - t _{0}}{\tau } \right]^{2/3}}
\exp{\left[- \frac{t - t _{0}}{\tau } \right]}\ ,
\label{55}
\end{eqnarray}
for the radiation energy density, respectively.
The dynamics of the entire system is determined by the energy densities 
(\ref{54}) and (\ref{55}) together with the field equations for
$a \left(t \right)$. In the case of a
spatially flat FLRW universe the latter reduce to the Friedmann
equation
\begin{equation}
3 \frac{\dot{a}^{2}}{a ^{2}} =
8 \pi \left(\rho _{_{\left(BH \right)}}
+ \rho _{_{\left(2 \right)}}\right)\ .
\label{56}
\end{equation}
We have numerically integrated the integro-differential set of
equations (\ref{54}), (\ref{55}) and (\ref{56}) and shown the results 
for $\rho _{_{\left(BH \right)}}$ and $\rho _{_{\left(2 \right)}}$
in figures 1 and 2, respectively. We have included the possibility  
of no black hole at all ($\beta = 0$)
for the sake of completeness and comparison with the
evolution of a uncontaminated FLRW radiation universe.
The impact of the evaporation process on the expansion rate may be  
understood in terms of an effective bulk viscous pressure of the  
cosmic fluid as a whole as will be discussed in the following  
sections.
It is well known that a bulk pressure tends to increase the expansion 
rate \cite{ZTP,ZGeq}.

Denoting the number of relativistic particles by
$N _{_{\left(2 \right)}} = n _{_{\left(2 \right)}}a ^{3}$,
the following general relations hold for the radiation component  
through the whole process:
\begin{equation}
\rho _{_{\left(2 \right)}} \propto
\frac{N _{_{\left(2 \right)}}^{4/3}}{a ^{4}}\ ,
\mbox{\ \ \ }
T _{_{\left(2 \right)}} \propto
\frac{N _{_{\left(2 \right)}}^{1/3}}{a } = n _{_{\left(2  
\right)}}^{1/3}\ ,
\label{57}
\end{equation}
implying $\rho _{_{\left(2 \right)}} \propto T _{_{\left(2  
\right)}}^{4}$.

Toward the end of the
evaporation, i.e. for $t \rightarrow t _{0} + \tau $, tha rates
$\dot{m}_{_{\left(BH \right)}}/m _{_{\left(BH \right)}}$ and
$\Gamma _{_{\left(2 \right)}}$ diverge, leading to explosive  
particle emission
in the final stages of the evaporation process.
As a result we have an increase in the radiation energy density,  
equivalent to a second phase of intense reheating, whereas the  
energy density in the PBH component falls down sharply.
(We mention that there have been suggestions in the literature  
according to which the PBHs do not evaporate completely but leave  
behind Planck-sized remnants \cite{BCL2,Carretal}.
If this were the case, the final reheating would be reduced.)
By direct inspection we confirm that there is no phase with a constant 
ratio between both energies densities (see the discussion preceding and following Eq. (\ref{53})).

\section{The effective one-component model}
In the present section we try to find an effective
one-component description for the system of black holes and radiation. 
Since our considerations are based on the assumption
that at some time $t _{0}$ the temperatures $T _{_{\left(BH  
\right)}}$ and
$T _{_{\left(2 \right)}}$ coincide,
it seems tempting to introduce an equilibrium temperature
for the system as a whole, following the lines of \cite{WZ1}.
The problem here is different, however, compared with a mixture
of two conventional fluids.
Once the evaporation has started, an approximate equilibrium
is no longer maintained, since there are no interactions
like collisional events between the components.
With their shrinking the black holes become hotter and hotter,
while the radiation temperature decreases with the expansion of the  
universe,   apart from phases of reheating.
Therefore, the concept of an equilibrium temperature of the system  
as a whole intuitively might appear rather formal, except perhaps in  
the initial stage of the process.
With the help of this quantity, however, a unified description of  
the cosmic matter will be possible which comprises both conventional  
fluid aspects and properties characterizing the BH component.

In a general two-fluid model close to equilibrium the Gibbs equation is 
\begin{equation}
T \mbox{d}s = \mbox{d} \frac{\rho}{n} + p \mbox{d} \frac{1}{n}
- \left(\mu _{_{\left(1 \right)}}
- \mu _{_{\left(2 \right)}}\right) \mbox{d} \frac{n _{_{\left(1  
\right)}}}{n}
{\mbox{ , }}
\label{58}
\end{equation}
where $s$ is the entropy per particle.
In the case of interest here
the temperature $T$ is the equilibrium temperature of the
system of a conventional fluid and a PBH component.
The temperatures $T _{_{\left(1 \right)}}$ and
$T _{_{\left(2 \right)}}$ of the previous
sections do not appear as variables in the present effective
one-temperature description.
An (approximate) equilibrium for the entire system is usually assumed 
to be established through the interactions between the
subsystems on the right-hand sides of Eqs. (\ref{6}) and (\ref{7}). 
This reasoning does not hold in the case in which one of the components
are evaporating black holes. 
Here, it is the circumstance that radiation is  
evaporated at the BH temperature which may constitute a situation  
which is similar to a thermal equilibrium between two fluids.
We assume that analogously to relations (\ref{10})
the cosmic medium as a whole is characterized by equations
of state
\begin{equation}
p = p\left(n, n _{_{\left(1 \right)}}, T\right) \ ,
\mbox{\ \ \ }
\rho = \rho \left(n, n _{_{\left(1 \right)}}, T\right){\mbox{ . }}
\label{59}
\end{equation}
If the first component is a black hole component,
the first of the equation of state (\ref{59}) specifies to
\begin{equation}
p = p _{2}\left(n _{2}, T\right)
\label{60}
\end{equation}
because of Eq. (\ref{22}).

If in the expressions
(\ref{18}) for the entropies per particle
the temperatures $T _{_{\left(1 \right)}}$ and
$T _{_{\left(2 \right)}}$ are identified among
themselves and with $T$, and $\rho \left(T\right) =
\rho _{_{\left(1 \right)}}\left(T\right) +
\rho _{_{\left(2 \right)}}\left(T\right)$ as well as  $p \left(T\right) 
= p _{_{\left(1 \right)}}\left(T\right)
+ p _{_{\left(2 \right)}}\left(T\right)$ are used,
the description based on Eq. (\ref{58}) is consistent with
the one relying on Eq. (\ref{8})  for $ns \left(T\right) =
n _{_{\left(1 \right)}} s _{_{\left(1 \right)}}\left(T\right)
+ n _{_{\left(2 \right)}} s _{_{\left(2 \right)}}\left(T\right)$.

The equilibrium temperature $T$ is {\it defined} by
\cite{UI}, \cite{ZMN}
\begin{equation}
\rho_{_{\left(1 \right)}}
\left(n_{_{\left(1 \right)}},T_{_{\left(1 \right)}}\right)
+ \rho_{_{\left(2 \right)}}
\left(n_{_{\left(2 \right)}},T_{_{\left(2 \right)}}\right)
= \rho \left(n, n _{_{\left(1 \right)}}, T\right)
{\mbox{ .}}
\label{61}
\end{equation}

As was shown by one of the authors \cite{WZ1}, nonvanishing source terms 
$\Gamma _{_{\left(A \right)}}$
in the particle number balances (\ref{4}) (deviations from
``detailed balance'')
give rise to a new
type of effective bulk pressure for the cosmic medium as a whole
(``reactive'' bulk pressure), i.e. to enlarged entropy production.  
Here we try to apply this concept to the case in which the process  
responsible for the deviations from ``detailed balance'' is PBH  
evaporation.

In order to find an explicit description we write
the energy-momentum tensor of the system
as a whole tentatively as \cite{WZ1}
\begin{equation}
T ^{ik} = \rho u ^{i}u ^{k} + \left(p + \Pi\right) h ^{ik}
{\mbox{ ,}}
\label{62}
\end{equation}
i.e., we try to map essential features of the evaporation process  
on an effective bulk pressure $\Pi $ which we determine below by  
consistency requirements.
The relations $T ^{ik}_{;k} = 0$
imply the energy balance
\begin{equation}
\dot{\rho} + \Theta \left(\rho + p + \Pi\right) = 0
\label{63}
\end{equation}
for the effective one-temperature description.

>From the Gibbs-equation (\ref{58}) one finds for the change
in the entropy per particle
\begin{eqnarray}
n \dot{s} &=& - \frac{\Theta }{T}\Pi  - \frac{\rho + p}{T}\Gamma  
\nonumber\\
&& \mbox{\ \ \ \ \ \ \ \ }- \frac{n _{_{\left(1 \right)}} n  
_{_{\left(2 \right)}}}{n}
\left(\frac{\mu _{_{\left(1 \right)}} - \mu _{_{\left(2  
\right)}}}{T}\right)
\left[\Gamma _{_{\left(1 \right)}} -
\Gamma _{_{\left(2 \right)}}\right] {\mbox{ .}}
\label{64}
\end{eqnarray}
The expression for the entropy
production density becomes
\begin{equation}
S ^{a}_{;a} = ns \Gamma + n \dot{s} {\mbox{ .}}
\label{65}
\end{equation}
Introducing here Eq. (\ref{64}) and the effective one-component  
chemical potential
\begin{equation}
\mu  = \frac{\rho + p}{n} - T s \ ,
\label{66}
\end{equation}
we find
\begin{eqnarray}
S ^{a}_{;a} &=& - \frac{\Theta}{T}\Pi  - \frac{n \mu}{T}\Gamma  
\nonumber\\
&&\mbox{\ \ \ }- \frac{n _{_{\left(1 \right)}} n _{_{\left(2  
\right)}}}{n}
\left(\frac{\mu _{_{\left(1 \right)}}
- \mu _{_{\left(2 \right)}}}{T}\right)
\left[\Gamma _{_{\left(1 \right)}}
- \Gamma _{_{\left(2 \right)}}\right] {\mbox{ . }}
\label{67}
\end{eqnarray}
Identifying again the first fluid with the black hole component,  
using Eqs.  (\ref{17}), (\ref{29}) and (\ref{36})  as well as the  
decomposition
\begin{equation}
n \mu  = n _{_{\left(BH \right)}} \mu _{_{\left(BH \right)}}
+ n _{_{\left(2 \right)}} \mu _{_{\left(2 \right)}}
= \frac{1}{2}\rho _{_{\left(BH \right)}}\ ,
\label{68}
\end{equation}
(recall that $\mu _{_{\left(2 \right)}} = 0$)
one obtains
\begin{equation}
S ^{a}_{;a} = - \frac{\Theta }{T}\Pi - \frac{1}{2}
\frac{\rho _{_{\left(BH \right)}}}{T}\Gamma _{_{\left(BH \right)}}\ .
\label{69}
\end{equation}
The last expression for the entropy production density has to be  
consistent with
Eq. (\ref{40}) of the two-component description.
This requirement provides us with the following expression for the  
effective viscous pressure,
\begin{eqnarray}
- \frac{\Theta }{T}\Pi &=& \frac{1}{2}\rho _{_{\left(BH \right)}}
\left(\Gamma _{_{\left(BH \right)}}
+ 2 \frac{\dot{m}_{_{\left(BH \right)}}}{m _{_{\left(BH \right)}}}\right)
\left[\frac{1}{T _{_{\left(BH \right)}}}
- \frac{1}{T _{_{\left(2 \right)}}}\right] \nonumber\\
&&+ \frac{1}{2}\rho _{_{\left(BH \right)}} \Gamma _{_{\left(BH \right)}}
\left[\frac{1}{T } - \frac{1}{T _{_{\left(2 \right)}}}\right]\ .
\label{70}
\end{eqnarray}
The last relation shows that
the evaporation process may be described in terms of  an effective  
viscous pressure $\Pi $.  In other words,
{\it  PBH evaporation, if regarded as a specific case of interfluid  
reactions,  gives rise to a viscous pressure of the cosmic medium  
as a whole.}
The viscous pressure vanishes for
$\Gamma _{_{\left(BH \right)}}
= \dot{m}_{_{\left(BH \right)}}/m _{_{\left(BH \right)}}
= \Gamma _{_{\left(2 \right)}} = 0$, i.e., no net change in the  
fluid particle numbers (``detailed balance'').
This is also the limiting case of noninteracting fluids.
We emphasize again that the quantity $\Pi $ in Eq. (\ref{70}) is  
exclusively a consequence of the evaporation process.  
``Conventional'' bulk viscous fluid pressures, e.g., due to  
scattering of radiation particles by the BHs, have been neglected  
here for simplicity.

By the field equations for a viscous cosmic medium with the  
energy-momentum tensor (\ref{62}) the quantity $\Pi $ influences the  
cosmic dynamics.
Restricting ourselves again to the homogeneous, isotropic and  
spatially flat case we have
\begin{equation}
3 \frac{\dot{a}^{2}}{a ^{2}} = 8 \pi  \rho \ ,
\mbox{\ \ }
\left(\frac{\dot{a}}{a} \right)^{\displaystyle \cdot}
= - 4 \pi \left(\rho + p + \Pi  \right)\ .
\label{71}
\end{equation}
While physically the present one-component picture is equivalent to  
the two-component description of the previous section, the mapping  
of certain features of the evaporation process onto an effective  
bulk pressure $\Pi $ may provide a more transparent understanding of  
how the PBH evaporation modifies
the cosmological dynamics.

In Eq. (\ref{70})
one has to deal with three generally different temperatures:
The temperatures $T _{_{\left(BH \right)}}$ and $T _{_{\left(2  
\right)}}$
and the equilibrium temperature $T$ of the system as a
whole, defined by Eq. (\ref{61}).
The dynamics of $T _{_{\left(BH \right)}}$ and
$T _{_{\left(2 \right)}}$ is given by the laws (\ref{32}) and  
(\ref{34}), respectively.
But we have still to find a corresponding relationship
for $T$.
Such a law may be obtained via similar steps that led
us to Eq. (\ref{12}).
There exists, however, the following complication.
Because of the additional dependence of $\rho $ on $n _{_{\left(1  
\right)}}$
one has now three partial derivatives of $\rho $:
$\left(\partial \rho / \partial T\right)_{n, n _{\left(1 \right)}}$, 
$\left(\partial \rho / \partial n\right)_{T, n _{\left(1 \right)}}$, and 
$\left(\partial \rho / \partial n _{_{\left(1 \right)}}\right)_{n, T}$. 
The requirement that $s$ is a state function now leads to
(\cite{ZiSpon}, \cite{ZCQG})
\begin{eqnarray}
\frac{\partial \rho }{\partial n} &=& \frac{\rho + p}{n}
- \frac{T}{n}\frac{\partial p}{\partial T} \nonumber\\
&&- \frac{n _{_{\left(1 \right)}}}{n}
\left[\left(\mu _{_{\left(1 \right)}} - \mu _{_{\left(2 \right)}}\right) 
- T \frac{\partial }{\partial T}
\left( \mu _{_{\left(1 \right)}} - \mu _{_{\left(2 \right)}}
\right)\right] {\mbox{ , }}
\label{72}
\end{eqnarray}
generalizing Eq. (\ref{11}),
and the additional relation
\begin{equation}
\frac{\partial \rho }{\partial n _{\left(1 \right)}} =
\mu _{_{\left(1 \right)}} - \mu _{_{\left(2 \right)}}
- T \frac{\partial }{\partial T}
\left( \mu _{_{\left(1 \right)}} - \mu _{_{\left(2 \right)}}\right) 
{\mbox{ .}}
\label{73}
\end{equation}
Using the Gibbs-Duhem relations
\begin{equation}
\mbox{d} p _{_{\left(A \right)}} = n _{_{\left(A \right)}}
s _{_{\left(A \right)}} \mbox{d} T _{_{\left(A \right)}}
+ n _{_{\left(A \right)}}
\mbox{d}\mu _{_{\left(A \right)}}
\label{74}
\end{equation}
for $T _{_{\left(A \right)}} = T$ together with Eq. (\ref{18}), one finds
\begin{equation}
\mu _{_{\left(A \right)}}
- T \frac{\partial \mu _{_{\left(A \right)}}}{\partial T}
= \frac{\rho _{_{\left(A \right)}}}
{n _{_{\left(A \right)}}}
{\mbox{ .}}
\label{75}
\end{equation}
Consequently, the relations (\ref{72}) and (\ref{73}) may be
written as
\begin{equation}
\frac{\partial \rho }{\partial n} = \frac{\rho + p}{n}
- \frac{T}{n}\frac{\partial p}{\partial T}
- \frac{n _{_{\left(1 \right)}}}{n}
\left[\frac{\rho _{_{\left(1 \right)}}}{n _{_{\left(1 \right)}}}
- \frac{\rho _{_{\left(2 \right)}}}{n _{_{\left(2 \right)}}} \right]
\label{76}
\end{equation}
and
\begin{equation}
\frac{\partial \rho }{\partial n _{_{\left(1 \right)}}} =
\frac{\rho _{_{\left(1 \right)}}}{n _{_{\left(1 \right)}}}
- \frac{\rho _{_{\left(2 \right)}}}{n _{_{\left(2 \right)}}}
{\mbox{ , }}
\label{77}
\end{equation}
respectively,
with $\rho _{_{\left(1 \right)}} = \rho _{_{\left(1  
\right)}}\left(T\right)$
and
$\rho _{_{\left(2 \right)}} = \rho _{_{\left(2  
\right)}}\left(T\right)$, since we are
within the one-temperature description.

Differentiating the second of the relations (\ref{59}) and applying  
Eqs. (\ref{4}), (\ref{17}),
(\ref{63}), (\ref{76}), and (\ref{77}),
we obtain
\begin{eqnarray}
\frac{\partial \rho }{\partial T} \dot{T} &=&
- \left(\Theta - \Gamma \right) T \frac{\partial p}
{\partial T}
- \left[\Theta \Pi  + \Gamma \left(\rho + p\right)\right] \nonumber\\
&& \mbox{\ \ \ } - \frac{n _{_{\left(1 \right)}}
n _{_{\left(2 \right)}}}{n}
\left(\frac{\rho _{_{\left(1 \right)}}}{n _{_{\left(1 \right)}}}
- \frac{\rho _{_{\left(2 \right)}}}{n _{_{\left(2 \right)}}}\right)
\left[\Gamma _{_{\left(1 \right)}}
- \Gamma _{_{\left(2 \right)}}\right]
{\mbox{ .}}
\label{78}
\end{eqnarray}
Using here $\rho = \rho _{_{\left(BH \right)}}
+ \rho _{_{\left(2 \right)}}$ and the second relation of Eqs.  
(\ref{17}), we find the evolution law for the equilibrium  
temperature of a system of a conventional fluid  and a ``fluid'' of  
primordial single-mass black holes,
\begin{eqnarray}
T \frac{\partial{\rho }}{\partial{T}}
\frac{\dot{T}}{T} &=& - T \left(\Theta - \Gamma _{_{\left(2  
\right)}}\right)
\frac{\partial{p}}{\partial{T}}
+ \rho _{_{\left(BH \right)}}
\frac{\dot{m}_{_{\left(BH \right)}}}{m _{_{\left(BH \right)}}}
- \Theta \Pi \nonumber\\
&&  + \frac{n _{_{\left(BH \right)}}}{n}
\left(T \frac{\partial{p}}{\partial{T}} - p \right)
\left(\Gamma _{_{\left(BH \right)}} - \Gamma _{_{\left(2  
\right)}}\right)\ .
\label{79}
\end{eqnarray}
{\it The temperature law (\ref{79}) comprises both the law  
(\ref{34}) for the fluid temperature and the BH temperature law  
(\ref{32}).}
This unifying feature may be considered as a justification of the  
equilibrium temperature concept for the cosmic medium.
For $\rho _{_{\left(BH \right)}} = 0$, Eq. (\ref{79}) reduces to to  
the expression  (\ref{34})
for radiation while for
$\rho _{_{\left(2 \right)}} = p _{_{\left(2 \right)}} = \Pi = 0$,  
equivalent to
$\rho = \rho _{_{\left(BH \right)}}$ and $T = T _{_{\left(BH  
\right)}}$, it coincides with the behavior (\ref{32}), equivalent to  
Eq. (\ref{28}).

Inserting here the expressions (\ref{45}) at $t = t _{0}$, (\ref{46}) 
and (\ref{47})  and taking into account  $\Pi \left(t _{0} \right)  
= 0$ for
$T _{_{\left(BH \right)}}\left(t _{0} \right)
= T _{_{\left(2 \right)}}\left(t _{0} \right) =
T \left(t _{0} \right)$ [cf. Eq. (\ref{70})], Eq. (\ref{79}) at $t  
= t _{0}$ reduces to the perfect fluid temperature law
\begin{equation}
\frac{\dot{T}}{T} = - \frac{{\partial p}}
{\partial{\rho }}\Theta \ ,
\mbox{\ \ \ \ \ \ }
\left(t = t _{0} \right)\ .
\label{80}
\end{equation}
This result again proves  the consistency of the concept of an  
equilibrium temperature $T$ of the system as a whole.

\section{Entropy production and viscous pressure at the
beginning of the evaporation phase}
The temperature laws (\ref{32}), (\ref{34})
and (\ref{79}) allow us to calculate the entropy production
(\ref{69}) and the viscous pressure (\ref{70})
at the beginning of the
black hole evaporation explicitly.
For the temperatures we have in linear order:
\begin{equation}
T \left(t\right) \approx T \left(t _{0}\right)
- \left(t - t _{0} \right) \Theta
T \frac{\partial{p} }{\partial{\rho }} \ ,
\label{81}
\end{equation}
\begin{equation}
T _{_{\left(2 \right)}}\left(t\right)
\approx T _{_{\left(2 \right)}} \left(t _{0}\right)
- \left(t - t _{0} \right)
\left(\Theta - \Gamma _{_{\left(2 \right)}}\right)
T \frac{\partial{p}_{_{\left(2 \right)}}}
{\partial{\rho }_{_{\left(2 \right)}}}
\label{82}
\end{equation}
and
\begin{equation}
T _{_{\left(BH \right)}}\left(t\right)
\approx T _{_{\left(BH \right)}} \left(t _{0}\right)
- \left(t - t _{0} \right) T _{_{\left(BH \right)}}
\frac{\dot{m}_{_{\left(BH \right)}}}{m_{_{\left(BH \right)}}}\ .
\label{83}
\end{equation}
Assuming
\begin{equation}
 T  \left(t _{0}\right) = T _{_{\left(2 \right)}} \left(t _{0}\right) 
= T _{_{\left(BH \right)}} \left(t _{0}\right) \ ,
\label{84}
\end{equation}
the following relations hold at $t = t _{0}$:
\begin{eqnarray}
\frac{\partial{\rho }}{\partial{T}}
= \frac{\rho _{_{\left(2 \right)}} - \rho _{_{\left(BH  
\right)}}}{T} &\ ,&
\mbox{\ \ }
\frac{\partial p }{\partial \rho }
= \frac{1}{3} \frac{\rho _{_{\left(2 \right)}}}
{\rho _{_{\left(2 \right)}} - \rho _{_{\left(BH \right)}}} \ ,\nonumber\\
\frac{\rho _{_{\left(BH \right)}}}{\partial \rho / \partial T}
&=& T \frac{\rho _{_{\left(BH \right)}}}
{\rho _{_{\left(2 \right)}} - \rho _{_{\left(BH \right)}}} \ ,
\label{85}
\end{eqnarray}
and
\begin{equation}
\frac{\partial p }{\partial \rho}
- \frac{\partial p _{_{\left(2 \right)}}}{\partial \rho _{_{\left(2  
\right)}}}
= \frac{1}{3} \frac{\rho _{_{\left(BH \right)}}}
{\rho _{_{\left(2 \right)}} - \rho _{_{\left(BH \right)}}} \ .
\label{86}
\end{equation}
Using the rates (\ref{45}) at $t = t _{0}$, (\ref{46}) and  
(\ref{47}), we obtain
\begin{equation}
T \approx \left[1 - \frac{t - t _{0}}{3 \tau }
\frac{1}{1 - \beta }\Theta \tau \right]T \left(t _{0} \right)\ ,
\label{87}
\end{equation}
\begin{equation}
T _{_{\left(2 \right)}}\approx \left[1 - \frac{t - t _{0}}{3 \tau }
\left(\Theta \tau  - \beta  \right)\right]T \left(t _{0} \right)\ ,
\label{88}
\end{equation}
and
\begin{equation}
T _{_{\left(BH \right)}}\approx \left[1 + \frac{t - t _{0}}{3 \tau }
\right]T \left(t _{0} \right)\ .
\label{89}
\end{equation}
Either from Eq. (\ref{40}) or Eqs. (\ref{69}) and (\ref{70}) we  
find for the entropy production density
\begin{equation}
S ^{a}_{;\ a} \approx \frac{1}{2 \tau }
\frac{\rho _{_{\left(BH \right)}}}{T}
\left[1 - \frac{t - t _{0}}{9 \tau }
\left(7 + 8 \beta + 3 H _{0} \tau    \right)   \right] \ .
\label{90}
\end{equation}

The back reaction of the evaporation process on the cosmological  
dynamics is
determined by the viscous pressure $\Pi $ which becomes in lowest  
order,
\begin{equation}
\Pi = - \frac{5 - 8 \beta }{1 - \beta }
\left[\frac{1 - \beta + 3 H _{0} \tau }{3 H _{0} \tau } \right]
\frac{t - t _{0}}{18 \tau }\rho _{_{\left(BH \right)}}\ .
\label{91}
\end{equation}
Except in the range $5/8 < \beta < 1$, the quantity $\Pi $ has a  
negative sign because of relation (\ref{49}).
Obviously, $\beta  \approx 1$ corresponding to
$\rho _{_{\left(2 \right)}}\left(t _{0}\right)
\approx \rho _{_{\left(BH \right)}}\left(t _{0}\right)$
is not a reasonable initial condition since $\Pi $ diverges.
The linear approximation breaks down for
$\rho _{_{\left(2 \right)}}\left(t _{0}\right)
\approx \rho _{_{\left(BH \right)}}\left(t _{0}\right)$.
As follows from the first relation (\ref{85}), the quantity
$\partial \rho / \partial T$ changes its sign for
$\rho _{_{\left(2 \right)}}\left(t _{0}\right)
= \rho _{_{\left(BH \right)}}\left(t _{0}\right)$.
A reasonable initial condition may be $\beta  = 1/2$, i.e., 
$\rho _{_{\left(2 \right)}}\left(t _{0}\right)
= 2 \rho _{_{\left(BH \right)}}\left(t _{0}\right)$.
In this case we find
\begin{equation}
\Pi = - \frac{1}{9}
\left[1 + \frac{1}{6 H _{0} \tau }\right]
\frac{t - t _{0}}{\tau }\rho _{_{\left(BH \right)}} \ ,
\mbox{\ \ \ }\left(\beta = \frac{1}{2}\right)\ .
\label{92}
\end{equation}
Since we restricted ourselves to the initial stage of the evaporation, 
$t - t _{0} < \tau $ is valid.
For $\beta  = 2$  equivalent to
$\rho _{_{\left(BH \right)}}\left(t _{0}\right)=
2\rho _{_{\left(2 \right)}}\left(t _{0}\right)$ one obtains
\begin{equation}
\Pi = - \frac{11}{18}
\left[1 - \frac{1}{ 3 H _{0} \tau }\right]
\frac{t - t _{0}}{ \tau }\rho _{_{\left(BH \right)}} \ ,
\mbox{\ \ \ }\left(\beta = 2\right)
\ .
\label{93}
\end{equation}
If the PBH component makes up almost all of the cosmic matter at
$t = t _{0}$ we have $\beta  \gg 1$ and
\begin{equation}
\Pi \approx - \frac{4}{9}
\left[1 - \frac{\beta }{3 H _{0} \tau }\right]
\frac{t - t _{0}}{ \tau }\rho _{_{\left(BH \right)}} \ ,
\mbox{\ \ \ }\left(\beta  \gg  1\right)\ .
\label{94}
\end{equation}
In the opposite limit $\beta  \ll 1$ the viscous pressure is
\begin{equation}
\Pi \approx - \frac{5}{18}
\left[1 + \frac{1}{3 H _{0} \tau }\right]
\frac{t - t _{0}}{ \tau }\rho _{_{\left(BH \right)}} \ ,
\mbox{\ \ \ }\left(\beta  \ll 1\right)\ .
\label{95}
\end{equation}
Although these explicit expressions for $\Pi $ are only valid in  
linear approximation, i.e. in the initial stage of the evaporation  
(we have $|\Pi| < \rho _{_{\left(BH \right)}}$ in all cases,  
consistent with our initial assumption of small deviations from  
equilibrium), they provide an idea of the direction of the process.
The role played by $\Pi $ in the field  equations (\ref{71})
may be thought of as a change of the ``$\gamma $-law''
( $\gamma = 1 + p/ \rho  $) of the cosmic fluid according to
\begin{equation}
\gamma \rightarrow \gamma \left[1 + \frac{\Pi }{\rho + p} \right]
\label{96}
\end{equation}
with
\begin{equation}
\frac{\Pi }{\rho + p} = - \frac{\beta }{4 + \beta }
\frac{5 - 8 \beta }{1 - \beta }
\left[\frac{1 - \beta + 3 H _{0} \tau }{3 H _{0} \tau } \right]
\frac{t - t _{0}}{6 \tau }\rho _{_{\left(BH \right)}}
\label{97}
\end{equation}
initially, i.e., $\gamma $ is effectively reduced.
Via the field equations, a negative viscous pressure tends to  
increase the expansion rate of the universe.

\section{Conclusions}
Based on the remarkable fact that the BH temperature law
$T _{_{\left(BH \right)}} \propto m _{_{\left(BH \right)}}^{-1}$
naturally fits into the general formula for the temperature of a  
fluid with variable particle number, we set up a scheme in which the  
evaporation of PBHs is interpreted as a deviation from ``detailed  
balance'' in an interacting and reacting two-fluid system.
A single-mass PBH configuration shares essential features with a  
pressureless, nonrelativistic fluid.
We modelled the evaporation process as a decay of such kind of  
``fluid'' into a conventional relativistic fluid (radiation).
We investigated the thermodynamics of this system, especially the  
entropy production and, with the help of the second law, derived  
general limits on the reheating of the radiation due to the  
evaporation.
We found that intense reheating, i.e., an increase in the radiation  
energy density, may occur both in the initial and final stages of  
the process.
For our single-mass PBH ensemble
there are no ``equilibrium'' solutions, i.e., solutions for which  
the ratio of the energy densities of the PBHs and the radiation  
remain constant,
as obtained in \cite{BCL} for a PBH configuration with a power-law  
mass spectrum.
The impact of the PBH evaporation on the cosmological dynamics may  
be described in terms of an effective bulk pressure of the cosmic  
substratum as a whole which we evaluated explicity for the initial  
phase of the process.
Its net effect is a higher expansion rate of the universe.

\ \\
{\bf Acknowledgement}\\
This paper was supported by the
Deutsche Forschungsgemeinschaft,
the Spanish Ministry of Education
(grant PB94-0718) and NATO
(grant CRG 940598).
We are grateful to Vicen\c{c} M\'{e}ndez for computational assistance.

\ \\

\ \\

\newpage
{\Large List of captions for figures}
\[\]
\underline{Figure 1}.- Evolution of the energy density of the
PBH fluid during the evaporation process.
\[\]
\underline{Figure 2}.- Evolution of the energy density of
the radiation fluid during the evaporation process.


\begin{thebibliography}{99}
\bibitem{ZEL} Ya. B. Zel'dovich and I. D. Novikov, Sov. Astron.
{\bf 10}, 602 (1967).
\bibitem{HAW1} S. W. Hawking, Mon. Not. R. Astr. Soc. {\bf 152},
75 (1971).
\bibitem{CHW} B. J. Carr and S. W. Hawking, Mon. Not. R.
Astr. Soc. {\bf 168}, 399 (1974).
\bibitem{CL} B. J. Carr and J. Lidsey, Phys. Rev. D {\bf 48},
543 (1993); {\bf 50}4853 (1994).
\bibitem{KP} M. Y. Khlopov and A. Polnarev, Phys. Lett. B
{\bf 97}, 383 (1980).
\bibitem{KoSaSa} H. Kodama, M. Sasaki, and K. Sato, 
Progr. Theor. Phys. {\bf 68}, 1979 (1982). 
\bibitem{BCKL} J.D. Barrow, E.J. Copeland, E.W. Kolb, and 
A.R. Liddle, Phys. Rev. D {\bf 43}, 984 (1991)
\bibitem{GPY} D. Gross, M. J. Perry, and L. G. Yaffe,
Phys. Rev. D {\bf 25}, 330 (1982); {\bf 36}, 1603 (1987).
\bibitem{KAP} J. I. Kapusta, Phys. Rev. D {\bf 30}, 831, (1984).
\bibitem{PW} T. Piran and R. M. Wald, Phys. Lett. {\bf 90} A, 20 (1982). 
\bibitem{RBH} R. Bousso and S. W. Hawking, Helv. Phys. Acta
{\bf 69}, 316 (1996); Phys. Rev. D {\bf 54}, 6312 (1996).
\bibitem{BLW} J. Garc\'{\i}a-Bellido, A. Linde, and D. Wands,
Phys. Rev. D {\bf 54}, 6040 (1996).
\bibitem{IVAN} P. Ivanov {\em Non-linear metric perturbations and
production of primordial black holes}, {\em preprint astro-ph/9708224
(1997)}.
\bibitem{PZ} A. Polnarev and R. Zemboricz, Phys. Rev. D {\bf 43},
1106 (1988).
\bibitem{JDB} J. D. Barrow, Mon. Not. R. Astr. Soc. {\bf 192}
427 (1980); D. Lindley, Mon. Not. R. Astr. Soc. {\bf 196}
317 (1981); L. M. Krauss, Phys. Rev. Lett {\bf 49}, 1459
(1982).
\bibitem{BJC} B. J. Carr, Helv. Phys. Acta {\bf 69}, 434 (1996).
\bibitem{BJC1} B. J. Carr, Astrophys. J. {\bf 201}, 1 (1975).
\bibitem{MG} M. Gibilisco, Int. J. Mod. Phys. A {\bf 11}, 5541
(1996).
\bibitem{GL} A. M. Green and A. R. Liddle,
Phys. Rev. D {\bf 56}, 6166 (1997).
\bibitem{GLR} A. M. Green, A. R. Liddle, and A. Riotto,
Phys. Rev. D {\bf 56}, 7559 (1997).
\bibitem{GD} G. Hayward and D. Pav\'on, Phys. Rev. D. {\bf 40},
1748 (1989).
\bibitem{BCL2} J.D. Barrow, E.J. Copeland, and A.R. Liddle,
Phys. Rev. D {\bf 46}, 645 (1992).
\bibitem{HAW2} S. W. Hawking, Commun. Math. Phys. {\bf 43},
199 (1975).
\bibitem{WZ1} W. Zimdahl, Mon. Not. R. Astr. Soc. {\bf 288},
665 (1997).
\bibitem{KCJ} K. C. Jacobs, Nature {\bf 215}, 1156 (1967).
\bibitem{BCL} J. D. Barrow, E. J. Copeland and A. R. Liddle,
Mon. Not. R. Astr. Soc., {\bf 253}, 675 (1991).
\bibitem{WD1} W. Zimdahl and D. Pav\'{o}n, Gen. Relativ. Grav.
{\bf 26}, 1259 (1994).
\bibitem{WD2} W. Zimdahl and D. Pav\'{o}n, Mon. Not. R. Astr. Soc.
{\bf 266}, 872 (1994).
\bibitem{WDR} W. Zimdahl, D. Pav\'{o}n, and R. Maartens,
Phys. Rev. D {\bf 55}, 4681 (1997).
\bibitem{Calv} M. O. Calv\~{a}o, J. A. S. Lima and I. Waga,
Phys. Lett. A {\bf 162}, 223 (1992); J. A. S. Lima and A. S. M.
Germano, Phys. Lett. A {\bf 170}, 373 (1992); W. Zimdahl and D.
Pav\'{o}n, Phys. Lett. A {\bf 176}, 57 (1993).
\bibitem{Groot} S.R. de Groot, W. A. van Leeuwen, and
Ch. G. van Weert, {\em Relativistic kinetic theory} (North-Holland,
Amsterdam, 1980).
\bibitem{Bern} J. Bernstein, {\em Kinetic theory in the expanding
universe} (Cambridge University Press, Cambridge, 1988).
\bibitem{Page1} D. N. Page, Phys. Rev. D {\bf 13}, 198 (1976).
\bibitem{TZP} K. S. Thorne, W. H. Zurek, and R. H. Price,
``The thermal atmosphere of a black hole" in
{\em  Black holes: The membrane paradigm}, edited by K. S. Thorne,
R. H. Price, and D. A. Macdonald (Yale University Press,
New Haven, 1986).
\bibitem{ZTP} W. Zimdahl, J. Triginer,
and D. Pav\'{o}n,
Phys. Rev. D {\bf 54}, 6101 (1996).
\bibitem{ZGeq} W. Zimdahl,
Phys. Rev. D {\bf 57}, 2245 (1998).
\bibitem{Carretal} B.J. Carr, J. Gilbert, and J. Lidsey,
Phys. Rev. D {\bf 50}, 4853 (1994).
\bibitem{UI} N. Udey and W. Israel, Mon. Not. R. Astr. Soc. {\bf 199},
1137 (1982).
\bibitem{ZMN} W. Zimdahl,
Mon. Not. R. Astr. Soc. {\bf 280},
1239 (1996).
\bibitem{ZiSpon} W. Zimdahl and H. Sponholz, Physica A {\bf 163},  
895 (1990).
\bibitem{ZCQG} W. Zimdahl, Class. Quantum Grav. {\bf 8}, 677 (1991).

\end{thebibliography}
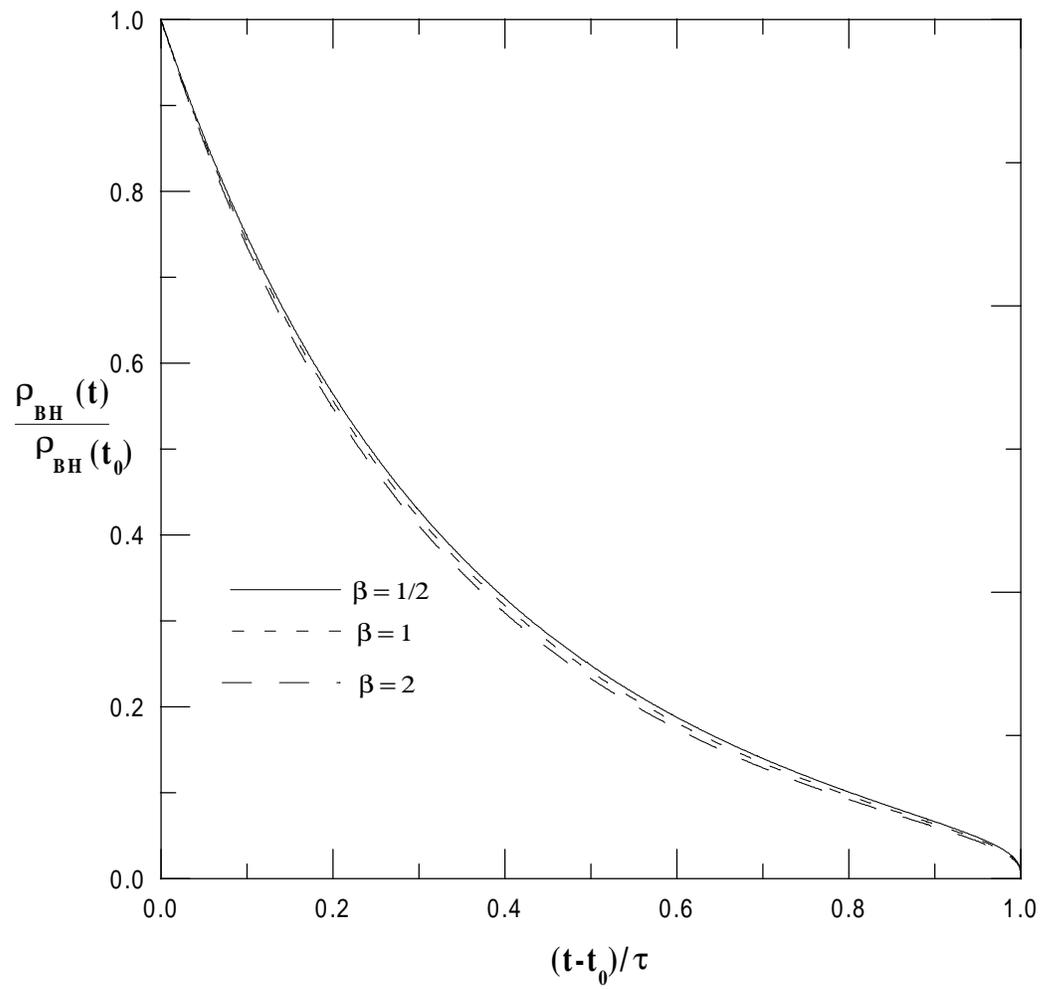
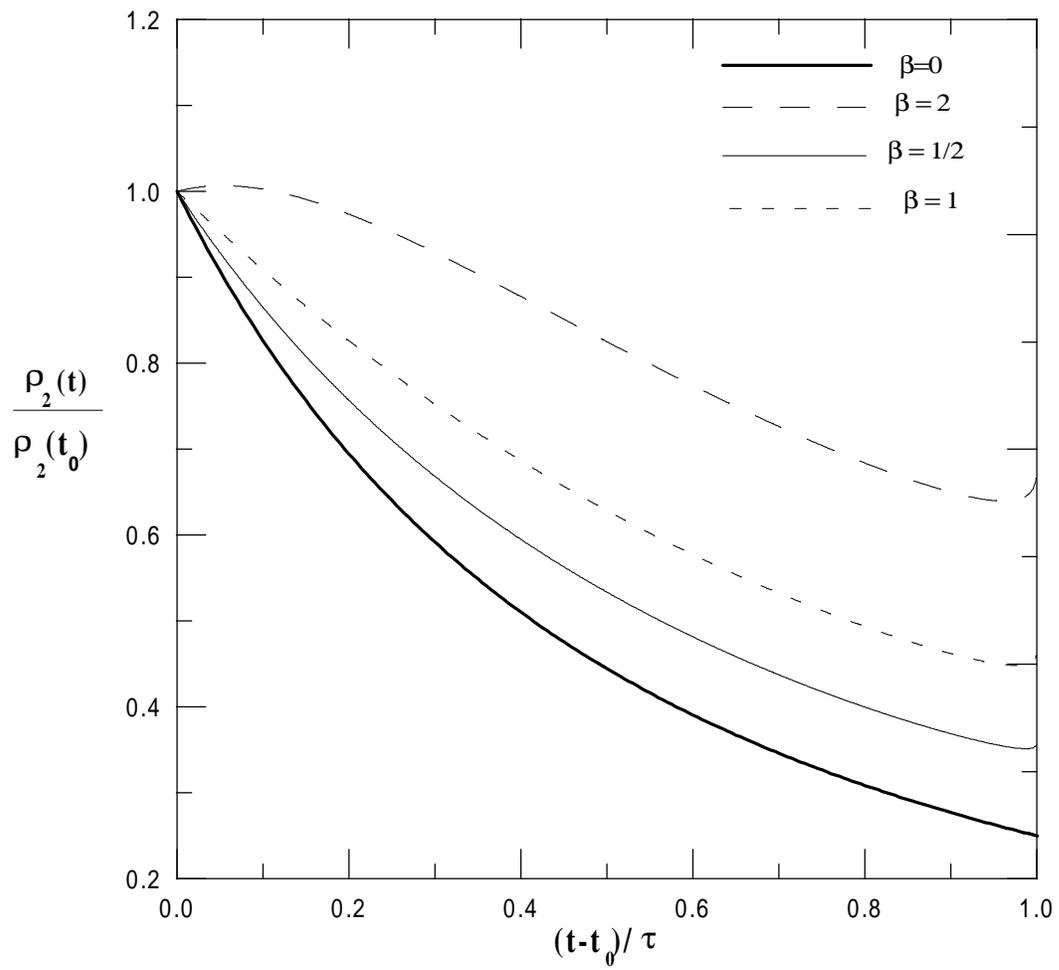
\end{document}